\documentstyle[11pt,AATS,epsf]{article}	

\begin{document}	

\title{Age Estimation of Extragalactic Globular Cluster Systems Using 
H$\beta$ Index} 

\author{Hyun-chul Lee, Suk-Jin Yoon, Young-Wook Lee}
\affil{Center for Space Astrophysics and Department of Astronomy, 
Yonsei University, 134 Shinchon, Seoul 120-749, Korea}


\begin{abstract}
After taking into account, for the first time, the detailed systematic 
variation of horizontal-branch (HB) morphology with age and 
metallicity, our population synthesis models show that the 
integrated H$\beta$ index is significantly affected by the presence of 
blue HB stars. Our models indicate that the strength of the
H$\beta$ index increases as much as 0.75 {\AA} due to blue HB stars. 
According to our models, a systematic difference between the globular 
cluster system in the Milky Way Galaxy and that in NGC 1399 in the 
H$\beta$ vs. Mg$_{2}$ diagram is understood if globular cluster 
systems in giant elliptical galaxies are couple of billion years older, 
in the mean, than the Galactic counterpart.
\end{abstract}


\section{Introduction}
For distant stellar populations, one relies upon the
integrated colors or spectra to investigate their ages and 
metallicities since individual stars are not resolved.
Here, we specifically focus on the H$\beta$ index, which
is widely used as an age indicator. Most of the previous works 
(e.g., Worthey 1994), however, have been done
on the basis that stars near the main-sequence turnoff
(MSTO) region are the most dominant sources for the integrated
strength of H$\beta$. Consequently, without meticulous consideration
for stars beyond the red giant branch, they claimed that 
the strength of H$\beta$ depends on the location
of the MSTO, which in turn depends on the age at a given metallicity.
Several investigators, however, have cast some doubt upon the 
sensitivity of the H$\beta$ index given the presence of other warm 
stars, especially blue horizontal-branch (HB) stars (see, e.g.,
Burstein et al. 1984; Jorgensen 1997).

On the observational side, it was barely possible to obtain
low signal-to-noise (S/N) spectra of globular clusters
in systems outside the Local Group (e.g., Mould et al. 1990). 
These spectra have been useful only for
kinematic information. With the advent of 10 m-class
telescopes, however, Kissler-Patig et al. (1998)
and Cohen, Blakeslee, \& Ryzhov (1998) have successfully obtained 
relatively high S/N spectra that provide reliable line index
calibration for globular clusters in NGC 1399 and M87,
the central giant elliptical galaxies in Fornax and Virgo clusters.

\section{Population Models with Horizontal-Branch Stars}
For our population models, the Yale Isochrones
(Demarque et al. 1996), rescaled for $\alpha$-element enhancement 
(Salaris, Chieffi, \& Straniero 1993), 
and the HB evolutionary tracks by Yi, Demarque, \& Kim (1997)
have been used. 
The Salpeter (1955) initial mass function is adopted
for the relative number of stars along the isochrones.
For the conversion from theoretical quantities to observable 
quantities, we have taken the most recently compiled stellar library 
of Lejeune, Cuisinier, \& Buser (1998) in order to cover
the largest possible ranges in stellar parameters
such as metallicity, temperature, and gravity.
The detailed calculation method of spectral index is presented 
in Lee, Yoon, \& Lee (2000).

In Figure 1a, the variations of H$\beta$ strength as a function of
metallicity (Mg$_{2}$) are plotted at given ages.
Here, $\Delta$t = 0 Gyr corresponds the recently favored
mean age of Galactic globular clusters ($\sim$ 12 Gyr) in the light of 
new distance scale as suggested by HIPPARCOS (e.g., Chaboyer et al. 
1998). Note that, unlike the models without HB stars
($dashed$ $lines$), distinct ``wavy" features 
appear in our models with HB stars ($solid$ $lines$). 
It is found that the strength of
H$\beta$ does not simply decrease with either increasing age or
increasing metallicity once HB stars are included in the models.
Now it is evident that the blue HB stars
around ($B$ $-$ $V$)$_{o}$ $\sim$ 0 are the key contributors
for the strength of H$\beta$. The differences in H$\beta$ strengths
between the models with and without HB stars are as much as 0.75 {\AA}
at the peak. 
In addition, it should be noted that the peak of the H$\beta$
enhancement moves to higher metallicity as age gets older.

\section{Comparison with Observations}
Having confirmed that the detailed modeling of HB is crucial
in the use of H$\beta$ index as an age indicator,
Figure 1b compares our results with observations of globular 
cluster systems in the Milky Way Galaxy and in NGC 1399. It is 
important to note here that all of these observations were carried out 
at the Keck telescope with the identical instrumental configuration, 
the Low Resolution Imaging Spectrograph (LRIS, Oke, de Zeeuw, \& Nemec 
1995). Despite the still large observational uncertainties,
it is inferred from Figure 1b that the NGC 1399 globular cluster system 
is perhaps systematically couple of billion years older,
in the mean, than the Galactic counterpart.

It is of considerable interest, in this respect, to find that a similar
age difference is inferred from the ``metal-poor HB solution" of the
UV upturn phenomenon of local giant elliptical galaxies (Park \& Lee 
1997; Yi et al. 1999). 
If our age estimation is confirmed to be correct,
this would indicate that the star formation in denser environments
has proceeded much more rapidly and efficiently, so that the
initial epoch of star formation in more massive (and denser) systems
occurred several billion years earlier than that of the Milky Way
(see also Lee 1992).


\acknowledgments
Hyun-chul Lee would like to thank the organizers, especially Ted, 
of this fantastic conference for their hospitality and for providing 
financial support.


\begin{references}
\reference Burstein, D., Faber, S. M., Gaskell, C. M., \& Krumm, N. 
    1984, \apj, 287, 586
\reference Chaboyer, B., Demarque, P., Kernan, P. J., \& Krauss, L. M. 
    1998, \apj, 494, 96
\reference Cohen J. G., Blakeslee, J. P., \& Ryzhov, A. 1998, \apj, 
    496, 808
\reference Demarque, P., Chaboyer, B., Guenther, D., Pinsonneault, M., 
    \& Yi, S.  1996, Yale Isochrone 1996
\reference Jorgensen, I. 1997, \mnras, 288, 161
\reference Kissler-Patig, M., Brodie, J. P., Schroder, L. L., Forbes, 
    D. A., Grillmair, C. J., \& Huchra, J. P. 1998, \aj, 115, 105
\reference Lee, H.-c., Yoon, S.-J., \& Lee, Y.-W. 2000, \aj, 120, 998
\reference Lee, Y.-W. 1992, \pasp, 104, 798
\reference Lejeune, T., Cuisinier, F., \& Buser, R. 1998, A\&AS, 130, 65
\reference Mould, J. R., Oke, J. B., de Zeeuw, P. T., \& Nemec, J. M. 
    1990, \aj, 99, 1823
\reference Oke, J. B., de Zeeuw, P. T., \& Nemec, J. M. 1995, \pasp, 
    99, 1823
\reference Park, J.-H., \& Lee, Y.-W. 1997, \apj, 476, 28
\reference Salaris, M., Chieffi, A., \& Straniero, O. 1993, \apj, 414, 
    580
\reference Salpeter, E. E. 1955 \apj, 121, 161
\reference Worthey, G. 1994, \apjs, 95, 107
\reference Yi, S., Demarque, P., \& Kim, Y.-C. 1997, \apj, 482, 677
\reference Yi, S., Lee, Y.-W., Woo, J.-H., Park, J.-H., Demarque, P., 
    \& Oemler, A., Jr. 1999, \apj, 513, 128
\end{references}
\end{document}